\documentclass{aastex63}

\received{}
\revised{}
\accepted{}

\submitjournal{AJ}

\shorttitle{Transients lasting less than a second}
\shortauthors{Arimatsu et al.}

\begin{document}

\title{Detectability of optical transients with timescales of sub-seconds}

\correspondingauthor{Ko Arimatsu}
\email{arimatsu@kwasan.kyoto-u.ac.jp}

\author{Ko Arimatsu}
\affiliation{The Hakubi Center/Astronomical Observatory, Graduate School of Science,  Kyoto University \\
Kitashirakawa-oiwake-cho, Sakyo-ku, \\
Kyoto 606-8502, Japan}

\author{Kohji Tsumura}
\affiliation{Department of Natural Science, Faculty of Science and Engineering, Tokyo City University, 
Setagaya, Tokyo, 158-8557, Japan}

\author{Fumihiko Usui}
\affiliation{Institute of Space and Astronautical Science, Japan Aerospace Exploration Agency,
3-1-1 Yoshinodai, Chuo-ku, Sagamihara, Kanagawa 252-5210, Japan}

\author{Takafumi Ootsubo}
\affiliation{Astronomy Data Center, National Astronomical Observatory of Japan
2-21-1 Osawa, Mitaka, Tokyo 181-8588, Japan}

\author{Jun-ichi Watanabe}
\affiliation{Public Relations Center, National Astronomical Observatory of Japan, 2-21-1 Osawa Mitaka Tokyo 181-8588 Japan}

\begin{abstract}
We search for optical transient sources with durations of $\sim 0.1$ to $\sim 1.3$~s using a dataset obtained in the Organized Autotelescopes for Serendipitous Event Survey (OASES) observation campaign. 
Since the OASES observations were carried out using two independent wide-field and high-cadence observation systems monitored the same field simultaneously, 
the obtained dataset provides a unique opportunity to develop a robust detection method for sub-second optical transients.
In the dataset of a selected field around the ecliptic and the Galactic plane,  
we find no astronomical event candidate that satisfies our detection criteria.  
From the non-detection result, we derive an upper limit on the event rate of sub-second transients 
around the ecliptic and the Galactic plane for the first time, 
obtaining $\sim 0.090$ and $\sim 0.38~{\rm hr^{-1}~deg^{-2}}$ for $m = 12$ and 13 Vmag, respectively.
In addition, future prospects of the sub-second scale transient event surveys are discussed.
\end{abstract}


\section{Introduction}
Optical astronomical transients with timescales of days to minutes have been recognized and studied over the past several centuries. 
On the other hand, 
there are only a few previous coordinated surveys exploring transients occurring over shorter, especially sub-second timescales.
For example, \citet{Schaefer87} present one of the earliest survey studies that explore 1--0.01~s scale optical transients using two telescopes equipped with a single photometer pointed at a random sky region.
They found 49 flash-like transient event candidates, 29 of which are those detected by both telescopes simultaneously.
These candidates are possibly caused by meteors and satellites passing through the observed sky region.
A survey concept of very short timescale transients is proposed by \citet{Griffin12} using photometer arrays onboard telescopes primarily designed for observations of optical Cherenkov flashes in the atmosphere.
\citet{Tingay20} demonstrates a drift scan technique for detections of optical transients with timescales much shorter than the integration time of CCD cameras.

Recently, observations with large-formatted and low-noise CMOS cameras provide opportunities for wide-field and unprecedented high-cadence observations. 
Early explorations of sub-second transients using CMOS cameras
have been carried out by the Mini-Mega-TORTORA survey \citep{Karpov16}, which uses nine independent small observation systems.
Each system consists of an 85-mm camera lens and a CMOS detector covering $10\degr \times 10\degr$.
In 2.5 years of the Mini-Mega-TORTORA survey, detections of optical transient candidates with durations of 0.4 to 1.0~s and typical peak magnitudes of 4 to 10 were reported \citep{Karpov17}. 
However, all of these candidate events are thought to be flashes caused by the reflection of sunlight from artificial satellites and space debris.
Recently, the Tomo-e gozen camera \citep{Sako18} consisting of 84 CMOS sensors covering $20~{\rm deg^2}$ in total and offering a 2-Hz sequential shooting mode was installed on the prime focus of the 105 cm Schmidt Telescope at Kiso Observatory.
Tomo-e gozen provides opportunities for not only high-cadence observations of known short-timescale astronomical events (e.g., \citealt{Arimatsu19b}) but also serendipitous surveys for unknown faint sources that emit only for a few seconds \citep{Richmond20}. 
However, the time resolution offered by Tomo-e gozen is slightly insufficient for explorations of optical transients with timescales comparable to or less than a second.

As of 2020, no clear evidence for non-repeating astronomical optical transients in the background with timescales less than $\sim1~{\rm s}$ has been reported. 
Furthermore, there is no observational constraint on the occurrence rate of these transient events. 
The lack of previous detections of short-timescale astronomical events is primarily due to insufficient optical surveys,
unlike previous surveys covering other wavelength ranges that led to historical discoveries of gamma-ray bursts (GRBs; \citealt{Klebesadel73}) and fast radio bursts (FRBs; \citealt{Lorimeret07}). 
This in turn means that there is a vast and unexplored parameter space of observational time-domain astronomy.
For example, \citet{Yang19} predict optical counterpart emissions from a small fraction of FRBs, 
which would reveal physical conditions in the FRB environments.
Recent superflare studies suggest that brightnesses of extremely powerful superflares on short timescales could reach approximately 100 times the quiescent emission \citep{Howard18}.
Thus the high-cadence monitoring of stellar fields should provide fruitful information on the occurrence rate of such extreme superflare activity.
Furthermore, especially in the sky close to the ecliptic, optical light flashes produced by mutual collisions of small asteroids would be observed like lunar impact flashes (e.g., \citealt{Yanagisawa02}).
Direct observations of these impact flashes should give a new insight into the production rate of interplanetary dust particles in the solar system.
Surveys with monitoring timescales less than a second are required to detect and investigate these possible transients.

Serendipitous surveys for optical sub-second transients face several challenges. 
First, one must observe with a cadence faster than 10 Hz to acquire multiple measurements of emitting events with timescales of less than a second. 
Second, one must monitor using instruments covering a large field of view (FOV) since these events are expected to be rare.
Third, one must achieve detections of non-repeating events that are robust concerning false-positive events, such as detector noises, 
atmospheric events including sporadic meteors,
and optical flashes due to the reflection of sunlight from artificial satellites and space debris. 
There are thus few previous wide-field monitoring observations for sub-second optical transients reported. 
Developing detection methods for sub-second transients and obtaining observational constraints for their occurrence rate should provide guidance to observations in the future.

This paper describes a survey for transient sources with durations of 0.2 to 2 s using a dataset obtained with two small observation systems named Organized Autotelescopes for Serendipitous Event Survey (OASES; \citealt{Arimatsu17}). 
Though the OASES project was originally carried out for detecting stellar occultations of small-sized trans-Neptunian objects \citep{Arimatsu19a}, 
they also provide a unique opportunity for understanding optical transients lasting less than a few seconds.
Based on the survey strategy described by \citet{Richmond20}, we searched the dataset obtained during an observation campaign in 2016-2017 with the OASES systems for short timescale transient events. 
The present results provide upper limits for astronomical transient events with timescales down to $\sim 0.1$~s, which is approximately an order of magnitude shorter than recent studies by \citet{Richmond20}.
Section 2 describes the outline of the datasets obtained by the OASES observations.
In section 3, a survey method developed for the present study including criteria of the transient detection and the survey results are presented. 
Upper limits on the occurrence rate of transient events derived from our survey results are presented in section 4.
Finally, we summarize the results and future prospects in section 5.

\section{Outline of the OASES Dataset}
The OASES monitoring observation campaign was carried out on a total of 23 nights between 25 June 2016 and 1 August 2017 UT.
The OASES project uses two identical observation systems (OASES-01 and OASES-02) primarily designed for detections of stellar occultation events by trans-Neptunian objects. 
Each system consists of a 279 mm Celestron $f = 2.2$ Rowe–Ackermann Schmidt Astrograph (RASA) equipped with a single ZWO ASI1600 MM-C CMOS camera and a Metabones Speed Booster SPEF-M43-BT4 focal reducer.
The effective focal ratio and angular pixel scale of the observation systems are $f/1.58$ and 1.96 arcsec, respectively. 
The OASES observation systems are capable of monitoring up to $\sim 2000$ stars with apparent brightnesses down to $V \sim 13.0$ in a $2\degr .3 \times 1\degr .8$ FOV simultaneously, providing signal-to-noise ratios comparable to or greater than $3 - 4$ with a sampling cadence of 15.4 Hz at an extremely low cost ($\sim 16000~{\rm USD}$ per single system).
Details of the OASES observation systems have been described in \citet{Arimatsu17}.

During the observation campaign, the two systems were installed in different positions on the rooftop of the Miyako open-air school ({\it Miyako Shonen Shizen no Ie}) on Miyako Island, Miyakojima-shi, Okinawa Prefecture, Japan, with a separation of 39~m (June 2016 to June 2017) or 52~m (July 2017 to August 2017).
For the monitoring observations, we selected a monitoring observation field with central equatorial, ecliptic, and galactic coordinates of $({\rm RA, Dec}) = (18:30:00, -22:30:00)$, $({\rm \lambda, \beta}) \sim (276 \degr .9, +0 \degr .8)$, and $(l, b) \sim (10 \degr .5, -5 \degr .6)$, respectively, to increase the detectability of stellar occultations of solar system objects. 
Since the selected field is close to the ecliptic and the galactic plane, transients occurring in the solar system's ecliptic plane and in the Galactic disk are expected to be detected in the OASES monitoring observations with higher sensitivities relative to randomly selected fields.
During the observations, images of the selected field are obtained with each observation system for a $2 \times 2$ binned sequential shooting mode of 15.4 frames per second. 
The exposure time is 65.0~ms for each frame. 
An individual single dataset consists of 3300 sequential frames.
The dataset obtained with the OASES monitoring observations in good weather conditions corresponds to 60~hrs of imaging data in total.
Details of the OASES monitoring observations have been described in \citet{Arimatsu19a}.

\section{Data reduction and transient detection method}
As noted by previous studies of high-cadence observations \citep{Karpov17, Richmond20}, 
optical flashes due to the reflection of sunlight from artificial satellites and space debris cause false detections of real astronomical transients.
To reduce these false detections, 
we have to use the data observed when the selected field was within the Earth's shadow.  
We thus select the image data of the selected field obtained when its Sun-observer-target angle was larger than $175\degr$ and Sun’s elevation above the horizon was lower than $-25\degr$.
The selected dataset is summarized in Table~1.
The total dataset that passes this criterion amounts to 10.1 hrs of data runs.

After dark subtraction, flat-fielding, and subtraction of a constant sky background level, 
stationary features (field stars, see Figure~1a and b) are subtracted and masked from each frame to produce {\it difference} images (Figure~1c and d).
As a reference frame, a median combined image is produced using 100 reduced images in every data group (Figure~1e and f).
To detect flash-like transients, we first run SExtractor (v2.5.0, \citealt{Bertin96}) on each difference image using a detection threshold of \texttt{DETECT\_THRESH} = 4.5.
We then manually run selection tests for detected point-like sources in the difference images according to the following criterion partially based on the previous second-scale transient survey by \citet{Richmond20}.

\begin{enumerate}
\item An event candidate must be detected between two and 20 times within a single data set of 3300 frames obtained with OASES-02.
This window size range corresponds to durations of $\sim 0.1$~s to $\sim 1.3$~s.
In the present study, we exclude single frame detections that can result in real transients with durations shorter than $\sim 0.1$~s.
We found that single-frame detection candidates are highly contaminated with false detections
such as short-timescale flashes of artificial satellites and space debris, detector noise spikes, and cosmic ray hitting.
Since the durations and the angular moving speeds of these candidates are unknown, 
it is difficult to select real astronomical event candidates from the contaminations.
Specific tests are thus required to select and investigate the single detections, which are beyond the scope of the present study.

\item All detections of a single event candidate must appear in a window of $(N + 2)$ consecutive frames, where $N$ is the total number of detections. 
In other words, events that appear once and then disappear once (i.e., non-repeating events) pass this criterion in general.
However, this criterion accepts up to two non-detection frames that are possibly caused by temporal variations in atmospheric transparency or scintillation effect.
\item An event candidate detected with OASES-02 must be detected more than one time simultaneously with OASES-01. 
The angular distance between the candidates observed with individual systems must be smaller than $4\arcsec$.
We note that the detection performance for faint objects of OASES-01 was slightly inferior to that of OASES-02 due to imperfect adjustment of its optical instrument.
\end{enumerate}

After running the selection tests above, we found three candidate events that satisfy these criteria.
An example of the candidate events is shown in Figure~1.
We investigate the central positions of each candidate on individual detected frames to obtain confidence that it is a real stationary object.  
All of these candidates are found to be moving objects at an angular speed of $5 - 10\arcsec~{\rm s}^{-1}$, which is consistent with that of artificial satellites or space debris in high-Earth orbit.
We have also investigated the angular parallaxes of the transient candidates 
by measuring their central positions obtained with the two systems at each time and found no significant displacement between the systems.
In conclusion, we found no clear candidate for astronomical transients with timescales less than $\sim 1$~s from the present dataset.
However, the detections of possible artificial objects demonstrate the capability of the present detection method to discover real transient events in a large amount of high-cadence dataset.

\section{Upper Limits of the Sub-second Astronomical Transients}
From the present non-detection result, we derived an upper limit of the occurrence rate of astronomical transient events.
The expected number of detected transient events $N_{\rm exp}(m)$ is expressed by:
\begin{eqnarray}
N_{\rm exp}(m) = \lambda(m)\,t\,\Omega(m),
\end{eqnarray}
where $\lambda(m)$, $t$, and $\Omega(m)$ are the occurrence rate of the transient event per unit solid angle with an apparent magnitude $m$, the total amount of observation data runs used for the present study (corresponding to $\sim 10.1~{\rm h}$),
and the effective angular survey area of the OASES observations, respectively. 
$\Omega(m)$ can be written as:
\begin{eqnarray}
\Omega(m) = \varepsilon(m)\,S,
\end{eqnarray}
where $\varepsilon(m)$ is the detection efficiency of a transient event with its apparent magnitude of $m$, 
and $S$ is the angular survey area of the OASES observation system ($4.05 \ {\rm deg^2}$).

We estimate $\varepsilon(m)$ by recovering artificial point sources implemented in randomly selected actual images obtained in the present dataset.
For this purpose, a point spread function for each image was constructed by stars detected in the same image.
The V-band apparent magnitudes of the artificial stars range from 11.75 to 13.0 with a 0.25 mag step.
For example, the estimated $\varepsilon(m)$ values for Vmag = 11.75, 12.5 and 13.0 are 0.87, 0.59 and 0.19, respectively.
Figure~2a shows the estimated $\Omega(m)$ for the present selected dataset as a function of $m$.

The upper limit to the transient event rate $\lambda(m)$ at 95\% confidence level as a function of $m$ is shown in Figure~2b and is also presented in Table~2.
$\lambda(m)$ is derived with a technique developed by \citet{Richmond20}, 
assuming a Poisson distribution of $N_{\rm exp}(m)$.
The obtained upper limit is placed approximately $0.1-0.4~{\rm hr^{-1} deg^{-2}}$ for the estimated magnitude range (Vmag$= 11.75-13.0$).
Since the selected field monitored in the present observations is close to the ecliptic and the Galactic plane,
the real occurrence rates of transients occurring around solar system's ecliptic plane and Galactic disk averaged over the entire sky are expected to be lower than the obtained upper limit.
On the other hand, typical amplitudes of the Galactic interstellar extinction toward the selected field are not significant but non-negligible ($A_V \sim 0.9$ mag; \citealt{Schlafly11}).
The occurrence rate of the extragalactic event could be higher than the obtained upper limit.

\section{Summary and Future Prospects}
With the dataset of a selected field around the ecliptic and the Galactic plane obtained with the OASES two observation systems simultaneously, 
a detection method for optical transient events with a timescale range of 0.1 to 1.3~s is developed. 
In the present dataset, we found no astronomical transient event candidates that satisfy our criteria.
From the non-detection results, an upper limit of the occurring rate of transient events was derived.
We should note that the present upper limit ($\sim 0.090$ and $\sim 0.38~{\rm hr^{-1}~deg^{-2}}$ for $m = 12$ and 13 Vmag, respectively) is larger than the occurrence rates of observable GRBs ($\sim 2\times 10^{-5}~{\rm hr^{-1} deg^{-2}}$; \citealt{vonKienlin20}) and that of FRBs ($\sim 10^{-2}~{\rm hr^{-1} deg^{-2}}$; \citealt{Thornton13}).
However, it is the first obtained observationally and provides guidance to observations in the future.

As already noted in Section 2, the previous OASES campaign only monitored the selected field close to the ecliptic and the Galactic plane. 
Future OASES campaigns plan to carry out monitoring of several fields to derive the ecliptic latitude and Galactic latitude dependences of the event rate.
As of 2020, a future upgrade of OASES observation systems is planned that will extend its current $2\degr.3  \times 1\degr.8 $ FOV
since the prime focus of the RASA optical tube offers a circular FOV with an angular diameter of $\sim 4 \degr$. 
Instead of the current front-illuminated sensor, a larger-sized back-illuminated CMOS sensor will be installed 
that offers a significantly larger (approximately $3\degr .3 \times 2\degr .5$) FOV and higher-sensitivity high-cadence imaging.
Monitoring observations of the upgraded OASES systems will provide an unprecedented opportunity of revealing faint and rare transient events with timescales of less than a second.

\clearpage

\begin{table}
  \caption{Summary of the OASES transient survey dataset}
  \label{tab1}
   \begin{center}
  \begin{tabular}{lll}
     \hline
    \hfil Date (UT) \hfil & hours  \\
    \hline
    2016 June 28  &  0.77   \\
    2016 June 29  &  0.95  \\
    2016 June 30  &  0.77  \\
    2016 July  02  &  1.61   \\
    2017 June 23  &  2.86   \\
    2017 June 27  &  0.71  \\
    2017 June 28   & 2.38   \\
       \hline
  \end{tabular}
    \end{center}
\end{table}

\begin{table}
  \caption{Upper limit of the occurrence rate $\lambda(m)$.}
  \label{tab2}
  \begin{center}
  \begin{tabular}{lll}
     \hline
    \hfil $m$ [Vmag] \hfil &  Upper lim. of $\lambda(m)$ ${\rm[hr^{-1}~deg^{-2}]}$ \\
    \hline
    11.75  &  0.083   \\
    12.0  &    0.090 \\
    12.25  &  0.10  \\
    12.5  &   0.13   \\
    12.75  & 0.19   \\
    13.0  &  0.38  \\
        \hline
  \end{tabular}
  \end{center}
\end{table}

\clearpage
\begin{figure}[!pt]
\begin{center}
   \includegraphics[scale=0.75]{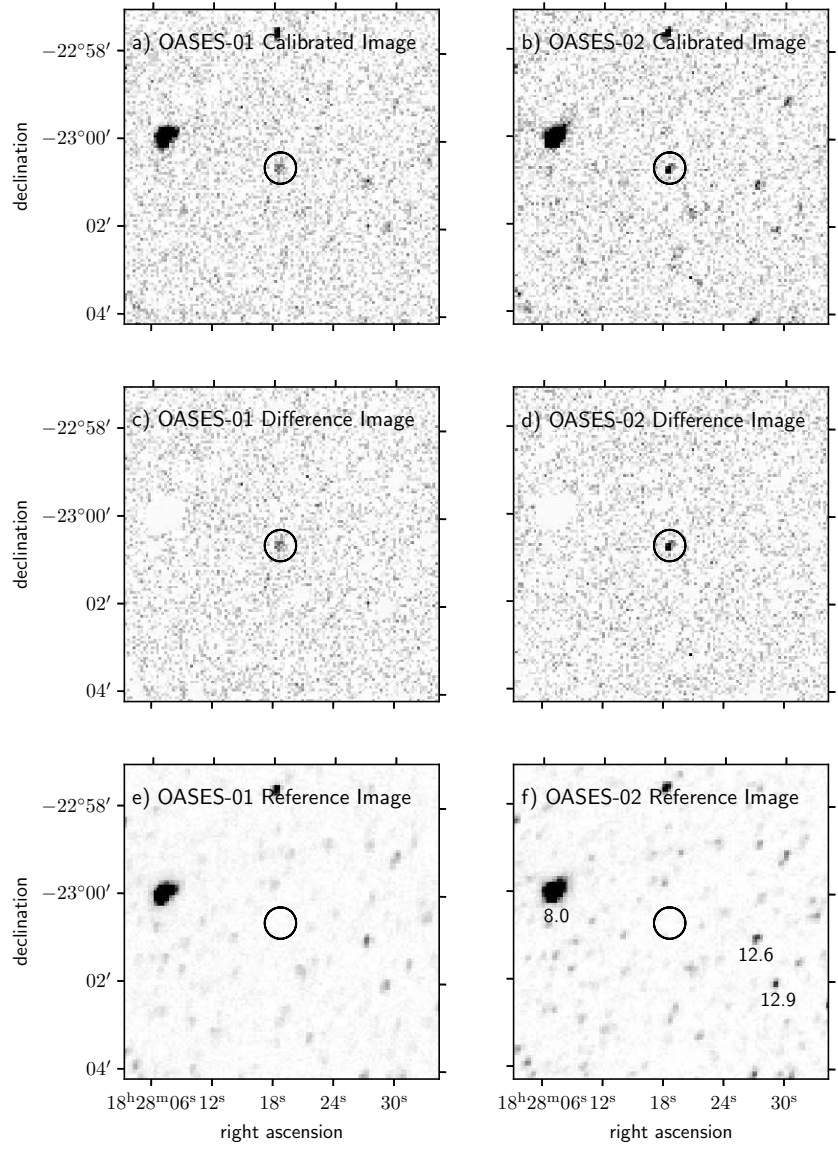}
   \caption{Example of transient event candidates (possibly a sunlight flash from a high-Earth orbit artificial object) detected through the present method.
   Panels (a) and (b) are calibrated images of a transient event candidate (shown in a black circle)
   obtained with OASES-01 and OASES-02, respectively.
   Panels (c) and (d) are the same as (a) and (b), but after the subtraction and masking of stationary features obtained by reference frames ((e) and (f), see Section~3), respectively.
   V-band magnitudes of field stars are also presented in panel (f).
  }
   \label{fig1}
 \end{center}
\end{figure}

\clearpage
\begin{figure}[!pt]
\begin{center}
   \includegraphics[scale=1]{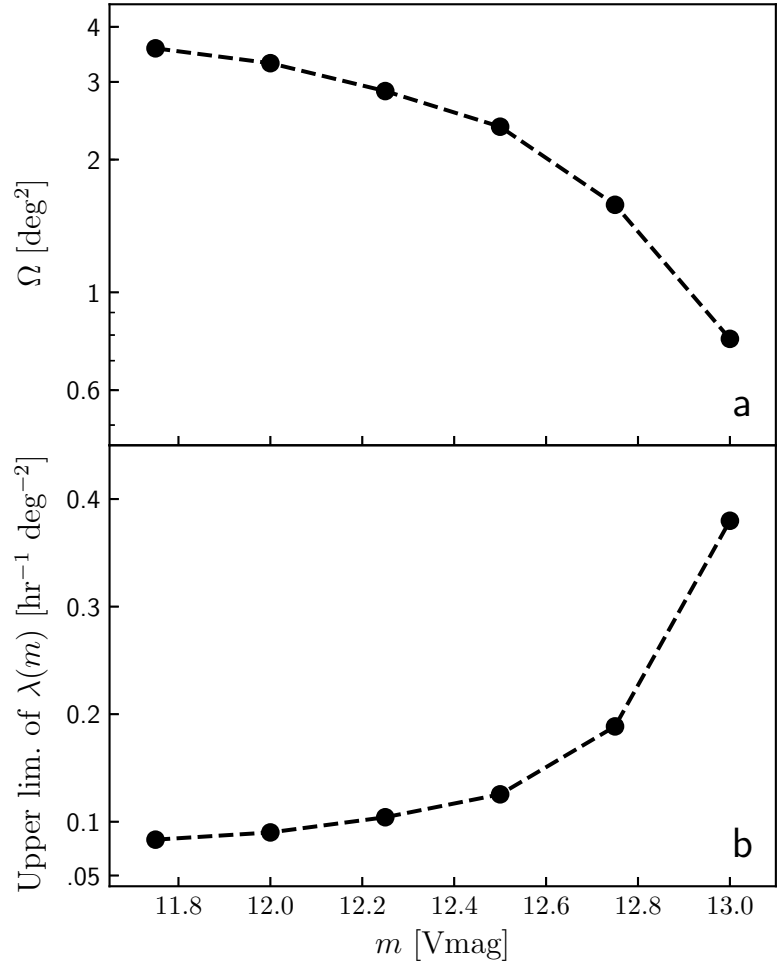}
   \caption{(a) Effective angular survey area $\Omega(m)$ for the present selected dataset as a function of V-band magnitude of transient sources $m$.
   (b) Upper limit of the occurrence rate of transient events $\lambda(m)$ as a function of $m$.
  }
   \label{fig2}
 \end{center}
\end{figure}

\clearpage

\acknowledgments
We thank an anonymous referee for a careful review and providing constructive suggestions.
This work has made use of data from the OASES observations partly carried out by Y. Shinnaka, K. Ichikawa, T. Kotani, T. Wada, K. Nagase, and Y. Sarugaku
with cooperation from Miyakojima City Museum, Miyako open-air school ({\it Miyako Shonen Shizen no Ie}), 
and people of the in Miyako Island.
This research has been partly supported by JSPS grants (16K17796, 18K13606).

\bibliography{sample63}{}
\bibliographystyle{aasjournal}

\end{document}